\documentclass[aps,prl,reprint,groupedaddress]{revtex4-2}
\usepackage[english]{babel}
\usepackage{graphicx}
\usepackage{float}
\usepackage{url}

\begin{document}


\title{Multicolor nonlinear chiral quantum optics: beyond phase}


\author{Cedric Dufresne}
\affiliation{Centre for Nanophotonics, Department of Physics,
Engineering Physics \& Astronomy, 64 Bader Lane,
Queen’s University, Kingston, Ontario, Canada K7L 3N6}
\email[Corresponding Author - Annabelle Makowski: ]{a.makowski@queensu.ca}
\author{Annabelle Makowski}
\affiliation{Centre for Nanophotonics, Department of Physics,
Engineering Physics \& Astronomy, 64 Bader Lane,
Queen’s University, Kingston, Ontario, Canada K7L 3N6}
\author{Nir Rotenberg}
\affiliation{Centre for Nanophotonics, Department of Physics,
Engineering Physics \& Astronomy, 64 Bader Lane,
Queen’s University, Kingston, Ontario, Canada K7L 3N6}


\begin{abstract}
Chiral quantum nonlinearities that arise when light interacts with quantum emitters are known to modulate only the phase but not the amplitude of scattered photons, enabling the creation of non-reciprocal photonic elements, quantum logic gates, and quantum network protocols. In this work, we show that the addition of a second photon beam drastically changes this picture, enabling both phase and amplitude modulation. Surprisingly, coherent photon transfer between the different beams enables a stronger amplitude modulation than  standard symmetric interactions. This is most obvious in the coherent, three-photon amplification, which we predict peaks with a $30\%$ efficiency in a chiral geometry, $3\times$ the efficiency of the symmetric configuration. Our results uncover a new regime of chiral quantum optics and provide a route towards more efficient all-optical control at few-photon energies.
\end{abstract}


\maketitle

\begin{figure*}
    \centering
    \includegraphics[]{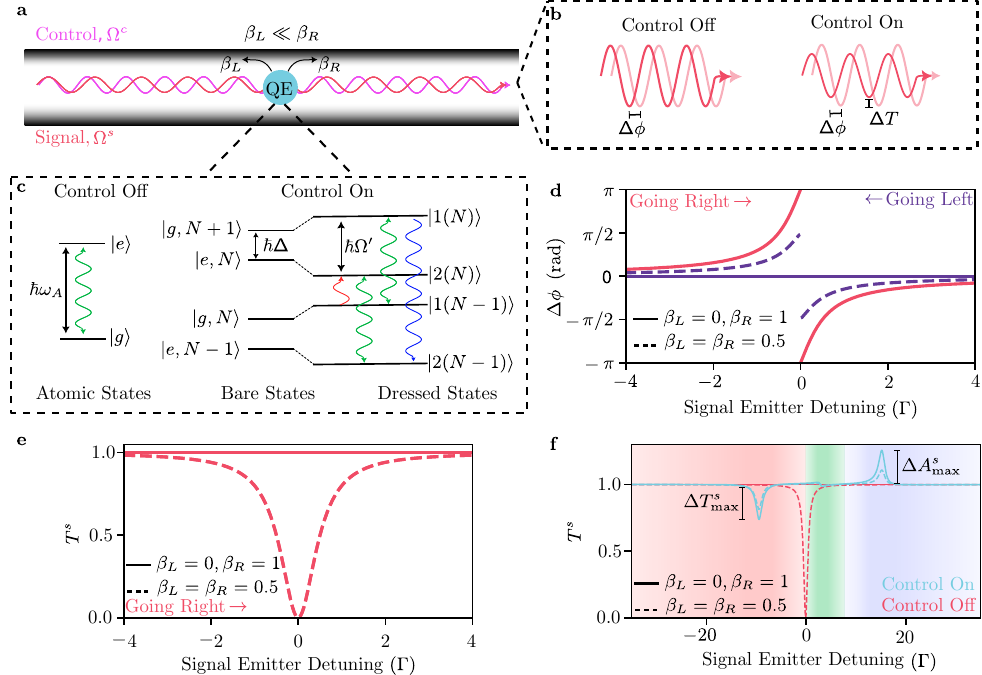}
    \caption{Multicolor quantum optics in a waveguide. a) A chirally coupled quantum emitter (QE) in a waveguide, here coupled efficiently to the right-propagating mode $\left(\beta_R \gg \beta_L\right)$ interacts with a weak signal beam and a control beam that may be much stronger. b) The signal is modulated by the interaction with the QE, which results in a phase shift, $\Delta\phi$, with no control beam and both a modulation of the phase and amplitude, $\Delta T$, when the control is on. c) This difference can be understood in terms of the modified energy landscape of the QE due to the control beam, the details of which depend on the control-emitter detuning $\Delta=\omega_c-\omega_A$ and the generalized Rabi frequency $\Omega'=\sqrt{4(\Omega^c)^2+\Delta^2}$ (\cite{c_cohen-tannoudji_dressed_1998} See Chapter 6 for details). The (d) phase-shift and (e) transmission of a right-propagating (red) and left-propagating (purple) signal interacting with a chirally-coupled (solid) and symmetric-coupled (dashed) QE with no control field. The presence of the chirally-coupled QE is only imprinted on the phase of the right propagating field (and never on the amplitude), while both the phase and amplitude of the field are altered due to interaction with a symmetric QE. (f) Exemplary spectra for the case when the control field is on (here, $\Omega^c = 6\Gamma, \Delta= 3\Gamma$), showing a strong modulation of amplitude for both the chiral and symmetric configurations.}
    \label{fig:idea}
\end{figure*}

The discovery of chiral quantum light-matter interactions, where the coupling of a quantum emitter to passing photons depends on the direction in which they travel, has been one of the great surprises of modern quantum optics \cite{lodahl_chiral_2017}. These chiral interactions arise when the efficiency with which a quantum emitter couples to photons traveling in the two counter-propagating directions of a waveguide differ; that is, $\beta_\mathrm{L} \neq \beta_\mathrm{R}$ shown in Fig~\ref{fig:idea}a. In contrast to a typical, symmetrically-coupled quantum emitter $\left(\beta_\mathrm{L} = \beta_\mathrm{R}\right)$, which in the ideal case acts like a perfect mirror, the transmission of a chirally-coupled emitter remains unity as seen in Fig.~\ref{fig:idea}e. Rather, the presence of the emitter is imprinted only on the phase of the scattered light that, ideally, is changed by $\pi$ when the photon (traveling in the correct direction) is on resonance with the emitter (Fig.~\ref{fig:idea}d). This hallmark of chiral quantum optics -- a large, directional phase shift -- has since been used to create nonreciprocal photonic devices such as an optical isolator \cite{kawaguchi_optical_2021,sayrin_nanophotonic_2015,tang_-chip_2019, xia_reversible_2014} and circulator \cite{scheucher_quantum_2016, wang_tunable_2020}, to demonstrate quantum logical gates \cite{wang_high-performance_2023,zhang_chirality_2022,zhou_chiral_2021} and spin-photon interfaces \cite{coles_chirality_2016,xiao_position-dependent_2021}, and been proposed as the basis for quantum photonic circuits \cite{mccaw_reconfigurable_2024,sollner_deterministic_2015,xiao_chiral_2021} and networks \cite{mahmoodian_quantum_2016,wang_chiral_2022,ramos_non-markovian_2016}.

Here, we show that in contrast to this accepted norm, chiral quantum interactions are capable of modulating both the photonic phase and amplitude. In contrast to the symmetric interactions, they do so without introducing reflections but rather modulate amplitude through a coherent nonlinear energy transfer. As we show, this occurs when the quantum emitter mediates a nonlinear interaction between different streams of photons, in the chiral analogue to Mollow's famous experiment \cite{wu_observation_1977}. The addition of the second photon stream, which we term the \textit{control} beam, dresses the bare emitter states, creating an infinite ladder of states that have both a light and a matter component. As a consequence, the energy landscape of the two-level emitter, with a single transition, is radically altered, resulting in three transitions \cite{c_cohen-tannoudji_dressed_1998}, as sketched in Fig.~\ref{fig:idea}c. Each of these transitions can be associated with a nonlinear effect: \textbf{(blue)} the AC Stark effect, which is a shift of spectral shift of the original resonance; \textbf{(green)} an energy transfer between the signal and control beams; \textbf{(red)} a coherent three-photon amplification of the signal. These, for a symmetrically-coupled emitter, have been observed using single organic molecules as emitters \cite{turschmann_chip-based_2017,maser_few-photon_2016} and, for the Stark shift, with a single quantum dot \cite{xu_single_2008}. An exemplary spectrum of the complex response of the dressed emitter is shown in Fig.~\ref{fig:idea}f, both for the symmetric (dashed) and chiral (solid) configurations. As we observe in this plot, and explain below, not only is the amplitude of the signal beam modulated in both cases, but for the correct choice of system parameters (detunings, powers) this modulation can be three times as large for the chiral geometry.

We begin by considering the signal transmission, $T^s$, which we write in terms of the signal field operators as~\cite{noauthor_see_nodate},
\begin{equation}\label{eq:Tsig}
    T^s = \frac{\left\langle\left(\hat{E}^{s-}_\mathrm{in} + \hat{E}^{s-}_\mathrm{sca}\right)\left(\hat{E}^{s+}_\mathrm{in} + \hat{E}^{s+}_\mathrm{sca}\right)\right\rangle}{\langle\hat{E}^{s-}_\mathrm{in}\hat{E}^{s+}_\mathrm{in}\rangle}.
\end{equation}
In this equation, $\hat{E}^{s\pm}_\mathrm{in}$ and $\hat{E}^{s\pm}_\mathrm{sca}$ are the positive (negative) frequency components of the incident and scattered signal fields, respectively. While the full derivation can be found in \cite{dufresne_place_nodate}, where we also consider the effect of all system imperfections, here we limit ourselves to the idealized case (no losses or noise and perfect directional coupling) where $\beta_\mathrm{R}=1$. This idealized scenario is sufficient to both uncover and understand the surprising physics of coherent chiral multicolor quantum nonlinearities.

Using a Green function formalism \cite{asenjo-garcia_atom-light_2017}, the total (positive frequency component) of the scattered field is,
\begin{equation}\label{eq:ESca}
   \hat{E}_{\mathrm{sca}}^{+} = i\Gamma \beta_{\mathrm{R}}\hat{\sigma}_{ge}e^{ikz},
\end{equation}
where $\Gamma$ is the emitter decay rate, $k$ the wavenumber to which the emitter couples, $z$ the position at which the field is measured, and $\hat{\sigma}_{ge}$ is the atomic operator corresponding to a transition from the excited to the ground state. In general, the scattered field will have components at the signal and control frequencies, $\omega^\mathrm{s}$ and $\omega^\mathrm{c}$, respectively, that beat in time. In Eq.~\ref{eq:ESca}, this time dependence is embedded in $\hat{\sigma}_{ge}$, which is not constant in time, even in the long-time limit under continuous excitation. We therefore calculate the Bloch equations using $\hat{\sigma}_{ge}$  \cite{jelezko_pumpprobe_1997}. In the case when the control beam is much stronger than the signal (i.e., $\Omega_\mathrm{c} \gg \Omega_\mathrm{s}$), we can then separate the time-averaged scattered \textit{signal} field from the total scattered field, allowing us to write the output signal field as~\cite{noauthor_see_nodate},
\begin{equation}\label{eq:Ts}
    T^{s} = 1 -\frac{2\beta_{R}\Gamma}{\Omega^s}\mathrm{Im}\{\rho_{eg,1}\} + \left(\frac{\beta_{R}\Gamma}{\Omega^{s}}\right)^2\left(\rho_{ee,0}-\rho_{ee}^{c}\right),
\end{equation}
where we use $\langle\hat{\sigma}_{ij}\rangle = \rho_{ji}$. We can do this because, in our Bloch expansion, $\rho_{eg,1}$ represents the steady-state coherence at the signal frequency, while $\rho_{ee,0}$ and $\rho_{ee}^{c}$ are the total and control-beam only steady-state emitter populations, respectively (where the latter is calculated by detuning the signal far from the emitter resonance \cite{noauthor_see_nodate}). Note that our approach fully separates the signal and control fields, including only photons due to the beam being considered, whereas earlier works include all other photons as a background \cite{turschmann_chip-based_2017,maser_few-photon_2016}.

\begin{figure}
    \centering
    \includegraphics[]{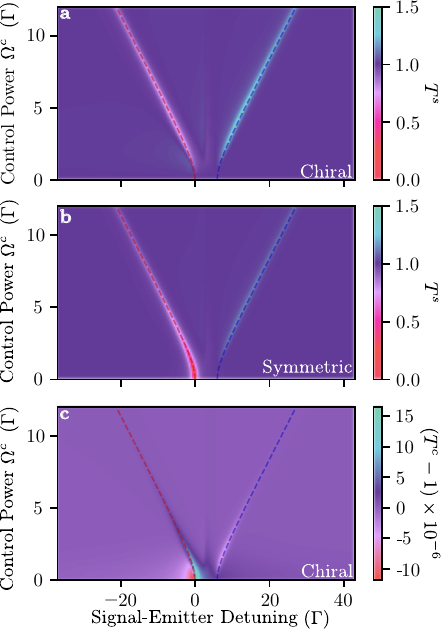}
    \caption{Control power-dependent transmission spectrum of the signal for (a) chiral and (b) symmetric scattering, and (c) of the control beam in the chiral geometry. In all cases, $\Delta = 3\Gamma$ and the signal-emitter detuning is scanned. In all panes, the transition corresponding to the AC Stark shift is denoted by a dashed red curve, while the three-photon amplification is shown by a dashed blue curve. As can be seen when comparing (a) to (c), photons lost to the signal are recovered in the control and vice versa, although the change to the control beam is about 5-orders of magnitude smaller (as this beam is much stronger than the signal $\left(\Omega^s=0.01\Gamma\right)$.}
    \label{fig:Ts}
\end{figure}

Equation~\ref{eq:Ts} allows us to calculate the transmitted signal, an example of which we show as a function of the signal-emitter detuning, $\delta_{se}$, and control beam power in Fig.~\ref{fig:Ts}(a) for a typical control-emitter detuning of $\Delta = 3\Gamma$. Surprisingly, with the control beam on $\left(\Omega^c \gtrsim \Gamma\right)$, signatures of all three dressed-state transitions (Fig.~\ref{fig:idea}b) are clearly imprinted on the signal amplitude. For this configuration, we observe a strong $\left(\sim 0.3\right)$ extinction when $\delta_{se}$ is negative and an equally strong amplification when $\delta_{se}$ is positive. This latter is especially surprising when considering the more typical symmetric geometry, shown in Fig.~\ref{fig:Ts}b, where the extinction is typically much stronger than the amplification, particularly for relatively low control powers. For example, when $\Omega^c = 3\Gamma$, the symmetric extinction peaks at $\Delta T^s_{\mathbf{max}=0.4}$, while the maximum amplification is only $\Delta A^s_{\mathbf{max}=0.1}$. In contrast, for the chiral interaction $\Delta T^s_{\mathbf{max}=0.35}$, slightly less than for the symmetric interactions, while $\Delta A^s_{\mathbf{max}=0.3}$. 

To understand the physics behind this striking difference, we consider the different channels available to signal photons after interacting with the emitter. In the ideal chiral geometry, the reflection is always zero (as confirmed by our model \cite{noauthor_see_nodate}, where we also confirm that energy is conserved) and only energy transfer between the signal and control are possible. Indeed, looking at the corresponding transmitted control spectra in Fig.~\ref{fig:Ts}c confirms that the extinction of $T^s$ for $\delta_{se}<0$ results in an amplification of $T^c$, and vice versa for positive $\delta_{se}$'s. In contrast, for the symmetric geometry, photons can also reflect, an effect that can enhance the extinction but competes with the amplification \cite{noauthor_see_nodate}. This suggests that the inherent nonlinearity of quantum emitters, as observed in effects such as three-photon amplification \cite{oelsner_dressed-state_2013,boyd_four-wave_1981}, is in fact larger than previously expected. And, it is only in chiral interactions that this full strength is revealed. 

\begin{figure}
    \centering
    \includegraphics[]{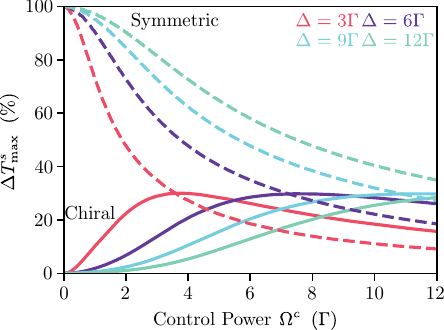}
    \caption{Maximum signal extinction for symmetric (dashed) and chiral (solid) coupling as a function of control power, for control-emitter detunings $\Delta = 3\Gamma, 6\Gamma, 9\Gamma$ and $12\Gamma$. The $\Delta = 3\Gamma$ curves (red), for example, follow the red, dashed lines in Fig.~\ref{fig:Ts}a and b.}
    \label{fig:nonLinear_Extinction}
\end{figure}

To more clearly visualize the nonlinear nature of these multicolor interactions we plot the control-power-dependent maximum extinction for both the chiral (solid curves) and symmetric (dashed curves) configurations, for different control-emitter detunings, in Fig.~\ref{fig:nonLinear_Extinction}. The red curves, for example, are for the case where $\Delta = 3\Gamma$, meaning that they show the $\Delta T^s$ values along the red dashed lines in Fig.~\ref{fig:Ts}a and b. Interestingly, the nonlinearity manifests in markedly different ways in each case. For the symmetric curve, the low-power extinction is unity, with the emitter acting like a perfect mirror as expected \cite{lodahl_chiral_2017}, decreasing monotonically as the power increases and decoheres the system. In contrast, for the chiral scenario there is no extinction initially, and it is only beyond $\Omega^c \approx \Gamma$ that the nonlinear response of the system is sufficient to shift photons from the signal to control beam. Consequently, the extinction due to the chiral interaction initially grows as the control power increases, only decreasing for high powers (above $\Omega^c \approx 4\Gamma$), albeit at a slower rate than in the symmetric configuration. The same overall power dependence is observed for larger $\Delta$'s, although an increase in control power is required to recover the same extinction due to the increasingly inefficient interaction between the control photons and the emitter.

\begin{figure}
    \centering
    \includegraphics[]{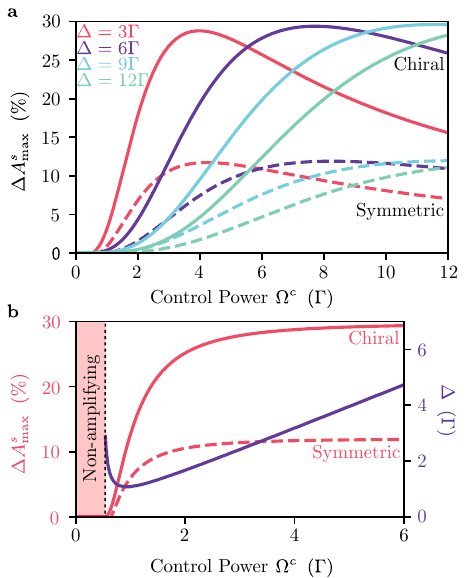}
    \caption{Three-photon amplification. (a) Maximum signal amplification for symmetric (dashed) and chiral (solid) coupling as a function of control power, for control-emitter detunings $\Delta = 3\Gamma, 6\Gamma, 9\Gamma$ and $12\Gamma$.  The $\Delta = 3\Gamma$ curves (red), for example, follow the blue, dashed lines in Fig.~\ref{fig:Ts}a and b. \textbf{b} Maximum achievable amplification (left axis, red curves) for each control power, for the symmetric and chiral configurations. For ease, points corresponding to curves shown in (a) are marked. The detuning, $\Delta$, at which the optimal amplification is found is shown in the purple curve (right axis).}
    \label{fig:amp}
\end{figure}

Similarly, the three-photon amplification depends on pump power in a highly nonlinear manner, as shown in Fig.~\ref{fig:amp}, although in a different way from the extinction. Following $\Delta A^s_{\mathrm{max}}$ for $\Delta = 3\Gamma$ (i.e., blue dashed curves in Figs.~\ref{fig:Ts}a and b), in the red curves in Fig.~\ref{fig:amp}a we observe that the amplification in both the chiral (solid) and symmetric (dashed) configuration increases once $\Omega^c$ increases beyond $\approx \Gamma$, peaks when $\Omega^c \approx 4\Gamma$, and then decreases beyond this value. We interpret this peak as arising due to a balance between the strength of the nonlinear response, power-induced decoherence, and emitter saturation, all of which increase with the control field. Strikingly, the chiral amplification is significantly stronger than that in the symmetric configuration, peaking at $\Delta A_{\mathrm{max}}^{\mathrm{sym}}= 11.8\%$ relative to  $\Delta A_{\mathrm{max}}^{\mathrm{chi}}= 28.9\%$ . As with the extinction, increasing the detuning simply necessitates a higher control power to recover this peak amplification, as the control-emitter interaction becomes less efficient. This is summarized in Fig.~\ref{fig:amp}b, where we plot $\Delta A^s_{\mathrm{max}}$ vs. $\Omega^c$, while also showing the $\Delta$ at which this maximum occurs (right axis). Here, we observe that the peak amplification quickly saturates, with the amplification in the chiral geometry almost 3 times larger than that calculated for a symmetrically-coupled emitter. 

In summary, we present a model for coherent multicolor chiral quantum optical nonlinearities and show how to separately calculate the scattered control and signal fields. In contrast to the accepted norm, we show that these chiral nonlinearities can strongly modulate the amplitude of the photonic wavefunction and not just its phase, and that this modulation is even stronger than for symmetric geometries \cite{turschmann_chip-based_2017,maser_few-photon_2016}. These results both illuminate the physics of non-degenerate few-photon quantum interactions, as well as provide potential routes towards future photonic technologies ranging from classical optical amplifiers for ultra-weak signals \cite{feng_ultra-low_2018,tong_ultralow_2012} to all-optical control of photonic wavefunctions for quantum applications \cite{jin_ultrafast_2014}.

\section{Acknowledgments}
The authors thank John E. Sipe for valuable discussions on the quantum theory, and gratefully acknowledge the support from the National Research Council of Canada (NRC), the Canadian Foundation for Innovation (CFI), the Ontario Ministry of Colleges, Universities, Research Excellence and Security (MCURES), the Natural Sciences and Engineering Research Council of Canada (NSERC), and Queen's University.

\bibliography{multicolor_nonlinear_chiral_quantum_optics_beyond_phase}

@article{jin_ultrafast_2014,
	title = {Ultrafast non-local control of spontaneous emission},
	volume = {9},
	copyright = {2014 Springer Nature Limited},
	issn = {1748-3395},
	url = {https://www.nature.com/articles/nnano.2014.190},
	doi = {10.1038/nnano.2014.190},
	number = {11},
	urldate = {2025-06-11},
	journal = {Nature Nanotechnology},
	author = {Jin, Chao-Yuan and Johne, Robert and Swinkels, Milo Y. and Hoang, Thang B. and Midolo, Leonardo and van Veldhoven, Peter J. and Fiore, Andrea},
	month = nov,
	year = {2014},
	pages = {886--890},
}

@article{jelezko_pumpprobe_1997,
	title = {Pump–probe spectroscopy and photophysical properties of single di-benzanthanthrene molecules in a naphthalene crystal},
	volume = {107},
	issn = {0021-9606},
	url = {https://doi.org/10.1063/1.474525},
	doi = {10.1063/1.474525},
	number = {6},
	urldate = {2025-05-27},
	journal = {The Journal of Chemical Physics},
	author = {Jelezko, F. and Lounis, B. and Orrit, M.},
	month = aug,
	year = {1997},
	pages = {1692--1702},
}

@article{kawaguchi_optical_2021,
	title = {Optical isolator based on chiral light-matter interactions in a ring resonator integrating a dichroic magneto-optical material},
	volume = {118},
	issn = {0003-6951},
	url = {https://doi.org/10.1063/5.0057558},
	doi = {10.1063/5.0057558},
	number = {24},
	urldate = {2025-05-28},
	journal = {Applied Physics Letters},
	author = {Kawaguchi, Yuma and Li, Mengyao and Chen, Kai and Menon, Vinod and Alù, Andrea and Khanikaev, Alexander B.},
	month = jun,
	year = {2021},
	pages = {241104},
}

@misc{noauthor_see_nodate,
	title = {See {Supplemental} {Material} [url] for [{Energy} {Conservation}, {Relfected} fields, {Control} background {Removal}]},
}

@article{tong_ultralow_2012,
	title = {Ultralow {Noise}, {Broadband} {Phase}-{Sensitive} {Optical} {Amplifiers}, and {Their} {Applications}},
	volume = {18},
	issn = {1558-4542},
	url = {https://ieeexplore.ieee.org/document/6096352},
	doi = {10.1109/JSTQE.2011.2136330},
	number = {2},
	urldate = {2025-05-28},
	journal = {IEEE Journal of Selected Topics in Quantum Electronics},
	author = {Tong, Zhi and Lundström, Carl and Andrekson, Peter A. and Karlsson, Magnus and Bogris, Adonis},
	month = mar,
	year = {2012},
	pages = {1016--1032},
}

@article{xiao_position-dependent_2021,
	title = {Position-dependent chiral coupling between single quantum dots and cross waveguides},
	volume = {118},
	issn = {0003-6951},
	url = {https://doi.org/10.1063/5.0042480},
	doi = {10.1063/5.0042480},
	number = {9},
	urldate = {2025-05-28},
	journal = {Applied Physics Letters},
	author = {Xiao, Shan and Wu, Shiyao and Xie, Xin and Yang, Jingnan and Wei, Wenqi and Shi, Shushu and Song, Feilong and Sun, Sibai and Dang, Jianchen and Yang, Longlong and Wang, Yunuan and Zuo, Zhanchun and Wang, Ting and Zhang, Jianjun and Xu, Xiulai},
	month = mar,
	year = {2021},
	pages = {091106},
}

@article{asenjo-garcia_atom-light_2017,
	title = {Atom-light interactions in quasi-one-dimensional nanostructures: {A} {Green}'s-function perspective},
	volume = {95},
	shorttitle = {Atom-light interactions in quasi-one-dimensional nanostructures},
	url = {https://link.aps.org/doi/10.1103/PhysRevA.95.033818},
	doi = {10.1103/PhysRevA.95.033818},
	number = {3},
	urldate = {2025-05-27},
	journal = {Physical Review A},
	author = {Asenjo-Garcia, A. and Hood, J. D. and Chang, D. E. and Kimble, H. J.},
	month = mar,
	year = {2017},
	pages = {033818},
}

@article{mahmoodian_quantum_2016,
	title = {Quantum {Networks} with {Chiral}-{Light}–{Matter} {Interaction} in {Waveguides}},
	volume = {117},
	url = {https://link.aps.org/doi/10.1103/PhysRevLett.117.240501},
	doi = {10.1103/PhysRevLett.117.240501},
	number = {24},
	urldate = {2025-05-28},
	journal = {Physical Review Letters},
	author = {Mahmoodian, Sahand and Lodahl, Peter and Sørensen, Anders S.},
	month = dec,
	year = {2016},
	pages = {240501},
}

@article{maser_few-photon_2016,
	title = {Few-photon coherent nonlinear optics with a single molecule},
	volume = {10},
	copyright = {2016 Springer Nature Limited},
	issn = {1749-4893},
	url = {https://www.nature.com/articles/nphoton.2016.63},
	doi = {10.1038/nphoton.2016.63},
	language = {english},
	number = {7},
	urldate = {2025-05-28},
	journal = {Nature Photonics},
	author = {Maser, Andreas and Gmeiner, Benjamin and Utikal, Tobias and Götzinger, Stephan and Sandoghdar, Vahid},
	month = jul,
	year = {2016},
	pages = {450--453},
}

@article{oelsner_dressed-state_2013,
	title = {Dressed-{State} {Amplification} by a {Single} {Superconducting} {Qubit}},
	volume = {110},
	url = {https://link.aps.org/doi/10.1103/PhysRevLett.110.053602},
	doi = {10.1103/PhysRevLett.110.053602},
	number = {5},
	urldate = {2025-05-28},
	journal = {Physical Review Letters},
	author = {Oelsner, G. and Macha, P. and Astafiev, O. V. and Il’ichev, E. and Grajcar, M. and Hübner, U. and Ivanov, B. I. and Neilinger, P. and Meyer, H.-G.},
	month = jan,
	year = {2013},
	pages = {053602},
}

@article{scheucher_quantum_2016,
	title = {Quantum optical circulator controlled by a single chirally coupled atom},
	volume = {354},
	url = {https://www.science.org/doi/10.1126/science.aaj2118},
	doi = {10.1126/science.aaj2118},
	number = {6319},
	urldate = {2025-05-28},
	journal = {Science},
	author = {Scheucher, Michael and Hilico, Adèle and Will, Elisa and Volz, Jürgen and Rauschenbeutel, Arno},
	month = dec,
	year = {2016},
	pages = {1577--1580},
}

@article{turschmann_chip-based_2017,
	title = {Chip-{Based} {All}-{Optical} {Control} of {Single} {Molecules} {Coherently} {Coupled} to a {Nanoguide}},
	volume = {17},
	issn = {1530-6984},
	url = {https://doi.org/10.1021/acs.nanolett.7b02033},
	doi = {10.1021/acs.nanolett.7b02033},
	number = {8},
	urldate = {2025-05-28},
	journal = {Nano Letters},
	author = {Türschmann, Pierre and Rotenberg, Nir and Renger, Jan and Harder, Irina and Lohse, Olga and Utikal, Tobias and Götzinger, Stephan and Sandoghdar, Vahid},
	month = aug,
	year = {2017},
	pages = {4941--4945},
}

@article{xu_single_2008,
	title = {Single {Charged} {Quantum} {Dot} in a {Strong} {Optical} {Field}: {Absorption}, {Gain}, and the ac-{Stark} {Effect}},
	volume = {101},
	shorttitle = {Single {Charged} {Quantum} {Dot} in a {Strong} {Optical} {Field}},
	doi = {10.1103/PhysRevLett.101.227401},
	number = {22},
	urldate = {2025-05-28},
	journal = {Physical Review Letters},
	author = {Xu, Xiaodong and Sun, Bo and Kim, Erik D. and Smirl, Katherine and Berman, P. R. and Steel, D. G. and Bracker, A. S. and Gammon, D. and Sham, L. J.},
	month = nov,
	year = {2008},
	pages = {227401},
}

@article{boyd_four-wave_1981,
	title = {Four-wave parametric interactions in a strongly driven two-level system},
	volume = {24},
	url = {https://link.aps.org/doi/10.1103/PhysRevA.24.411},
	doi = {10.1103/PhysRevA.24.411},
	number = {1},
	urldate = {2025-05-28},
	journal = {Physical Review A},
	author = {Boyd, Robert W. and Raymer, Michael G. and Narum, Paul and Harter, Donald J.},
	month = jul,
	year = {1981},
	pages = {411--423},
}

@article{wu_observation_1977,
	title = {Observation of {Amplification} in a {Strongly} {Driven} {Two}-{Level} {Atomic} {System} at {Optical} {Frequencies}},
	volume = {38},
	url = {https://link.aps.org/doi/10.1103/PhysRevLett.38.1077},
	doi = {10.1103/PhysRevLett.38.1077},
	number = {19},
	urldate = {2025-05-27},
	journal = {Physical Review Letters},
	author = {Wu, F. Y. and Ezekiel, S. and Ducloy, M. and Mollow, B. R.},
	month = may,
	year = {1977},
	pages = {1077--1080},
}

@article{wang_high-performance_2023,
	title = {High-performance chiral all-optical {OR} logic gate based on topological edge states of valley photonic crystal},
	volume = {32},
	issn = {1674-1056},
	url = {https://dx.doi.org/10.1088/1674-1056/accb41},
	doi = {10.1088/1674-1056/accb41},
	number = {7},
	urldate = {2025-05-28},
	journal = {Chinese Physics B},
	author = {Wang, Xiaorong and Fei, Hongming and Lin, Han and Wu, Min and Kang, Lijuan and Zhang, Mingda and Liu, Xin and Yang, Yibiao and Xiao, Liantuan},
	month = aug,
	year = {2023},
	pages = {074205},
}

@article{coles_chirality_2016,
	title = {Chirality of nanophotonic waveguide with embedded quantum emitter for unidirectional spin transfer},
	volume = {7},
	copyright = {2016 The Author(s)},
	issn = {2041-1723},
	url = {https://www.nature.com/articles/ncomms11183},
	doi = {10.1038/ncomms11183},
	number = {1},
	urldate = {2025-05-28},
	journal = {Nature Communications},
	author = {Coles, R. J. and Price, D. M. and Dixon, J. E. and Royall, B. and Clarke, E. and Kok, P. and Skolnick, M. S. and Fox, A. M. and Makhonin, M. N.},
	month = mar,
	year = {2016},
	pages = {11183},
}

@incollection{c_cohen-tannoudji_dressed_1998,
	title = {The {Dressed} {Atom} {Approach}},
	isbn = {978-3-527-61719-7},
	urldate = {2025-05-27},
	booktitle = {Atom—{Photon} {Interactions}},
	publisher = {John Wiley \& Sons, Ltd},
	author = {{C. Cohen-Tannoudji} and {D.-R. Jacques} and {G. Grynberg}},
	year = {1998},
	pages = {407--514},
}

@article{feng_ultra-low_2018,
	title = {Ultra-low noise optical injection locking amplifier with {AOM}-based coherent detection scheme},
	volume = {8},
	copyright = {2018 The Author(s)},
	issn = {2045-2322},
	url = {https://www.nature.com/articles/s41598-018-31381-x},
	doi = {10.1038/s41598-018-31381-x},
	number = {1},
	urldate = {2025-05-28},
	journal = {Scientific Reports},
	author = {Feng, Zitong and Yang, Fei and Zhang, Xi and Chen, Dijun and Wei, Fang and Cheng, Nan and Sun, Yanguang and Gui, Youzhen and Cai, Haiwen},
	month = sep,
	year = {2018},
	pages = {13135},
}

@article{sayrin_nanophotonic_2015,
	title = {Nanophotonic {Optical} {Isolator} {Controlled} by the {Internal} {State} of {Cold} {Atoms}},
	volume = {5},
	url = {https://link.aps.org/doi/10.1103/PhysRevX.5.041036},
	doi = {10.1103/PhysRevX.5.041036},
	number = {4},
	urldate = {2025-06-05},
	journal = {Physical Review X},
	author = {Sayrin, Clément and Junge, Christian and Mitsch, Rudolf and Albrecht, Bernhard and O’Shea, Danny and Schneeweiss, Philipp and Volz, Jürgen and Rauschenbeutel, Arno},
	month = dec,
	year = {2015},
	pages = {041036},
}

@article{tang_-chip_2019,
	title = {On-chip chiral single-photon interface: {Isolation} and unidirectional emission},
	volume = {99},
	shorttitle = {On-chip chiral single-photon interface},
	url = {https://link.aps.org/doi/10.1103/PhysRevA.99.043833},
	doi = {10.1103/PhysRevA.99.043833},
	number = {4},
	urldate = {2025-06-05},
	journal = {Physical Review A},
	author = {Tang, Lei and Tang, Jiangshan and Zhang, Weidong and Lu, Guowei and Zhang, Han and Zhang, Yong and Xia, Keyu and Xiao, Min},
	month = apr,
	year = {2019},
	pages = {043833},
}

@article{mccaw_reconfigurable_2024,
	title = {Reconfigurable quantum photonic circuits based on quantum dots},
	volume = {13},
	copyright = {De Gruyter expressly reserves the right to use all content for commercial text and data mining within the meaning of Section 44b of the German Copyright Act.},
	issn = {2192-8614},
	url = {https://www.degruyterbrill.com/document/doi/10.1515/nanoph-2024-0044/html},
	doi = {10.1515/nanoph-2024-0044},
	number = {16},
	urldate = {2025-05-27},
	journal = {Nanophotonics},
	author = {McCaw, Adam and Ewaniuk, Jacob and Shastri, Bhavin J. and Rotenberg, Nir},
	month = jul,
	year = {2024},
	pages = {2951--2959},
}

@article{wang_tunable_2020,
	title = {Tunable single-photon diode and circulator via chiral waveguide–emitter couplings},
	volume = {17},
	issn = {1612-202X},
	url = {https://dx.doi.org/10.1088/1612-202X/ab8557},
	doi = {10.1088/1612-202X/ab8557},
	number = {6},
	urldate = {2025-06-05},
	journal = {Laser Physics Letters},
	author = {Wang, Xin and Shui, Tao and Li, Ling and Li, Xiyun and Wu, Zhen and Yang, Wen-Xing},
	month = may,
	year = {2020},
	pages = {065201},
}

@unpublished{dufresne_place_nodate,
	title = {Chiral Nonlinear Optics and Optical Control},
	author = {Dufresne, Cedric and Makowski, Annabelle and Rotenberg, Nir},
	note = {(unpublished)}
}

@article{xia_reversible_2014,
	title = {Reversible nonmagnetic single-photon isolation using unbalanced quantum coupling},
	volume = {90},
	url = {https://link.aps.org/doi/10.1103/PhysRevA.90.043802},
	doi = {10.1103/PhysRevA.90.043802},
	number = {4},
	urldate = {2025-06-05},
	journal = {Physical Review A},
	author = {Xia, Keyu and Lu, Guowei and Lin, Gongwei and Cheng, Yuqing and Niu, Yueping and Gong, Shangqing and Twamley, Jason},
	month = oct,
	year = {2014},
	pages = {043802},
}

@article{sollner_deterministic_2015,
	title = {Deterministic photon–emitter coupling in chiral photonic circuits},
	volume = {10},
	copyright = {2015 Springer Nature Limited},
	issn = {1748-3395},
	url = {https://www.nature.com/articles/nnano.2015.159},
	doi = {10.1038/nnano.2015.159},
	number = {9},
	urldate = {2025-06-05},
	journal = {Nature Nanotechnology},
	author = {Söllner, Immo and Mahmoodian, Sahand and Hansen, Sofie Lindskov and Midolo, Leonardo and Javadi, Alisa and Kiršanskė, Gabija and Pregnolato, Tommaso and El-Ella, Haitham and Lee, Eun Hye and Song, Jin Dong and Stobbe, Søren and Lodahl, Peter},
	month = sep,
	year = {2015},
	pages = {775--778},
}

@article{zhang_chirality_2022,
	title = {Chirality logic gates},
	volume = {8},
	number = {49},
	urldate = {2025-06-10},
	journal = {Science Advances},
	author = {Zhang, Yi and Wang, Yadong and Dai, Yunyun and Bai, Xueyin and Hu, Xuerong and Du, Luojun and Hu, Hai and Yang, Xiaoxia and Li, Diao and Dai, Qing and Hasan, Tawfique and Sun, Zhipei},
	month = dec,
	year = {2022},
}

@article{zhou_chiral_2021,
	title = {Chiral single-photon switch-assisted quantum logic gate with a nitrogen-vacancy center in a hybrid system},
	volume = {9},
	copyright = {© 2021 Chinese Laser Press},
	issn = {2327-9125},
	url = {https://opg.optica.org/prj/abstract.cfm?uri=prj-9-3-405},
	doi = {10.1364/PRJ.405246},
	number = {3},
	urldate = {2025-06-10},
	journal = {Photonics Research},
	author = {Zhou, Yuan and Lü, Dong-Yan and Zeng, Wei-You},
	month = mar,
	year = {2021},
	pages = {405--415},
}

@article{xiao_chiral_2021,
	title = {Chiral {Photonic} {Circuits} for {Deterministic} {Spin} {Transfer}},
	volume = {15},
	copyright = {© 2021 Wiley-VCH GmbH},
	issn = {1863-8899},
	url = {https://onlinelibrary.wiley.com/doi/abs/10.1002/lpor.202100009},
	doi = {10.1002/lpor.202100009},
	number = {9},
	urldate = {2025-06-11},
	journal = {Laser \& Photonics Reviews},
	author = {Xiao, Shan and Wu, Shiyao and Xie, Xin and Yang, Jingnan and Wei, Wenqi and Shi, Shushu and Song, Feilong and Dang, Jianchen and Sun, Sibai and Yang, Longlong and Wang, Yunuan and Yan, Sai and Zuo, Zhanchun and Wang, Ting and Zhang, Jianjun and Jin, Kuijuan and Xu, Xiulai},
	year = {2021},
	pages = {2100009},
}

@article{lodahl_chiral_2017,
	title = {Chiral quantum optics},
	volume = {541},
	copyright = {2017 Macmillan Publishers Limited, part of Springer Nature. All rights reserved.},
	issn = {1476-4687},
	url = {https://www.nature.com/articles/nature21037},
	doi = {10.1038/nature21037},
	number = {7638},
	urldate = {2025-05-27},
	journal = {Nature},
	author = {Lodahl, Peter and Mahmoodian, Sahand and Stobbe, Søren and Rauschenbeutel, Arno and Schneeweiss, Philipp and Volz, Jürgen and Pichler, Hannes and Zoller, Peter},
	month = jan,
	year = {2017},
	pages = {473--480},
}

@article{wang_chiral_2022,
	title = {Chiral {Quantum} {Network} with {Giant} {Atoms}},
	volume = {7},
	issn = {2058-9565},
	doi = {10.1088/2058-9565/ac6a04},
	number = {3},
	urldate = {2026-01-21},
	journal = {Quantum Science and Technology},
	author = {Wang, Xin and Li, Hong-rong},
	month = jul,
	year = {2022},
	pages = {035007},
}

@article{ramos_non-markovian_2016,
	title = {Non-{Markovian} dynamics in chiral quantum networks with spins and photons},
	volume = {93},
	url = {https://link.aps.org/doi/10.1103/PhysRevA.93.062104},
	doi = {10.1103/PhysRevA.93.062104},
	number = {6},
	urldate = {2026-01-19},
	journal = {Physical Review A},
	author = {Ramos, Tomás and Vermersch, Benoît and Hauke, Philipp and Pichler, Hannes and Zoller, Peter},
	month = jun,
	year = {2016},
	pages = {062104},
}

\end{document}